\documentclass[aps,prb,twocolumn,superscriptaddress]{revtex4-1}
\usepackage{graphicx}
\usepackage{epstopdf}
\usepackage{amsmath}
\usepackage{amssymb}
\usepackage{natbib}
\usepackage[latin1]{inputenc}
\usepackage{enumerate}
\usepackage{dashrule}
\usepackage{mathtools}
\usepackage{color,soul}

\begin{document}
	
\title{Suppression of scattering in quantum confined 2D-helical Dirac systems}

\author{J. Dufouleur}
\email[email: ]{j.dufouleur@ifw-dresden.de}
\affiliation{Leibniz Institute for Solid State and Materials Research IFW Dresden, P.O. Box 270116, D-01171 Dresden, Germany}
\affiliation{Center for Transport and Devices, TU Dresden, D-01069 Dresden, Germany}

\author{E. Xypakis}
\affiliation{Max-Planck-Institut für Physik Komplexer Systeme, D-01187 Dresden, Germany}

\author{B. Büchner}
\affiliation{Leibniz Institute for Solid State and Materials Research IFW Dresden, P.O. Box 270116, D-01171 Dresden, Germany}
\affiliation{Department of Physics, TU Dresden, D-01062 Dresden, Germany}

\author{R. Giraud}
\affiliation{IFW Dresden, P.O. Box 270116, D-01171 Dresden, Germany}
\affiliation{Univ. Grenoble Alpes, CNRS, CEA, Grenoble-INP, INAC-Spintec, 38000 Grenoble, France}

\author{J. H. Bardarson}
\affiliation{Max-Planck-Institut für Physik Komplexer Systeme, D-01187 Dresden, Germany}
\affiliation{Department of Physics, KTH Royal Institute of Technology, Stockholm, SE-106 91 Sweden}

\date{\today}

\begin{abstract}
Transport properties of helical Dirac fermions in disordered quantum wires are investigated in the large energy limit. In the quasi-ballistic regime, the conductance and the Fano factor are sensitive to disorder only when the Fermi energy is close to an opening of a transverse mode. In the limit of a large number of transverse modes, transport properties are insensitive to the geometry of the nanowire or the nature and strength of the disorder but, instead, are dominated by the properties of the interface between the ohmic contact and the nanowire. In the case of a heavily doped Dirac metallic contact, the conductance is proportional to the energy with an average transmission $\mathcal{T}=\pi/4$ and a Fano factor of $F\simeq 0.13$. Those results can be generalized to a much broader class of contacts, the exact values of $\mathcal{T}$ and $F$ depending on the model used for the contacts. The energy dependence of Aharonov-Bohm oscillations is determined, revealing a damped oscillatory behavior and phase shifts due to the 1D-subband quantization and which are not the signature of the non-trivial topology.
\end{abstract}

\maketitle

Transport properties of two dimensional helical Dirac fermions were first studied in carbon nanotubes \cite{Ando1997, Ando1998, Ando1998a, McEuen1999, Bachtold1999} and more recently in graphene \cite{CastroNeto2009,DasSarma2011} and topological insulators \cite{Bardarson2013}. For massless fermions, the linear dispersion relation and the symmetries that constrain scattering not to connect orthogonal (pseudo)spins, induce a strongly anisotropic scattering, leading to an enhanced transport scattering time\cite{Hwang2008, Culcer2010, Monteverde2010, Dufouleur2016}. The long transport length $\ell$ results in large mobilities and in quantum confinement effects in disordered systems with one or more dimensions smaller than $\ell$. Prior work~\cite{CastroNeto2009,DasSarma2011} highlighted for instance the properties of Dirac fermions either in absence of quantum confinement \cite{Hwang2008, Culcer2010} or disorder \cite{Tworzydlo2006,Prada2007,Beenakker2009} or in presence of both but focusing on low-energy limit \cite{Ando1997, Ando1998, Ando1998a, Bardarson2010,Zhang2010a,Bardarson2013}. Although the transport properties are rather well understood close to the Dirac point, the coexistence of strong quantum confinement and high Fermi energy in Dirac systems is most common in real nanostructures, like 3D topological insulator nanowires \cite{Hong2014,Cho2015,Ziegler2017} or narrow graphene nanoribbons \cite{Lin2008,Tombros2011,Terres2016}.

A clear understanding of the interplay between scattering and quantum confinement far away from the Dirac point would shed light on some recent experimental works \cite{Peng2010,Xiu2011,Dufouleur2013,Hong2014,Cho2015,Jauregui2016,Dufouleur2017}. In this work, considering both metals contacts and disorder allows us to evaluate their relative contributions to the transport properties. Moreover, in the case of weakly disordered topological insulator nanowires, we calculate the energy dependence of Aharonov-Bohm oscillations close to and far from the Dirac point. This makes possible to distinguish between topologically trivial signatures (due to quantum confinement) and nontrivial features\cite{Hong2014,Cho2015,Jauregui2016}.

In the model considered here, we investigate the properties of disordered topological insulator nanowires with $\ell/W > 1$ where  $W$ is the perimeter of the nanowire. We consider the case of a perfect interface with metallic electrodes and we focus on the high-Fermi-energy regime ($\varepsilon \gg \Delta$ with $\Delta=hv/W$ and $v$ the Fermi velocity) in the presence of a magnetic field parallel to the nanowire axis. Starting from an approximate analytical derivation of the transmission of transverse modes, we calculate the transport properties at any energy and magnetic flux, including quantum corrections induced by intermode scattering. The comparison to numerical simulations validate our analytical approach and an excellent agreement is found at high energy as long as  $\ell/W > 1$. In this regime, the energy and disorder strength dependence of the conductance and shot noise reveal the ballistic nature of the transport. Furthemore, we evidence the damped oscillatory behavior in the energy dependence of Aharonov-Bohm oscillations away from the Dirac point and phase shifts. We show that it is related to quantum confinement only, in good quantitative agreement with experimental results \cite{Hong2014,Cho2015,Jauregui2016}.

\section{Model}

\begin{figure}[t]
	\centering
	\includegraphics[width=\columnwidth]{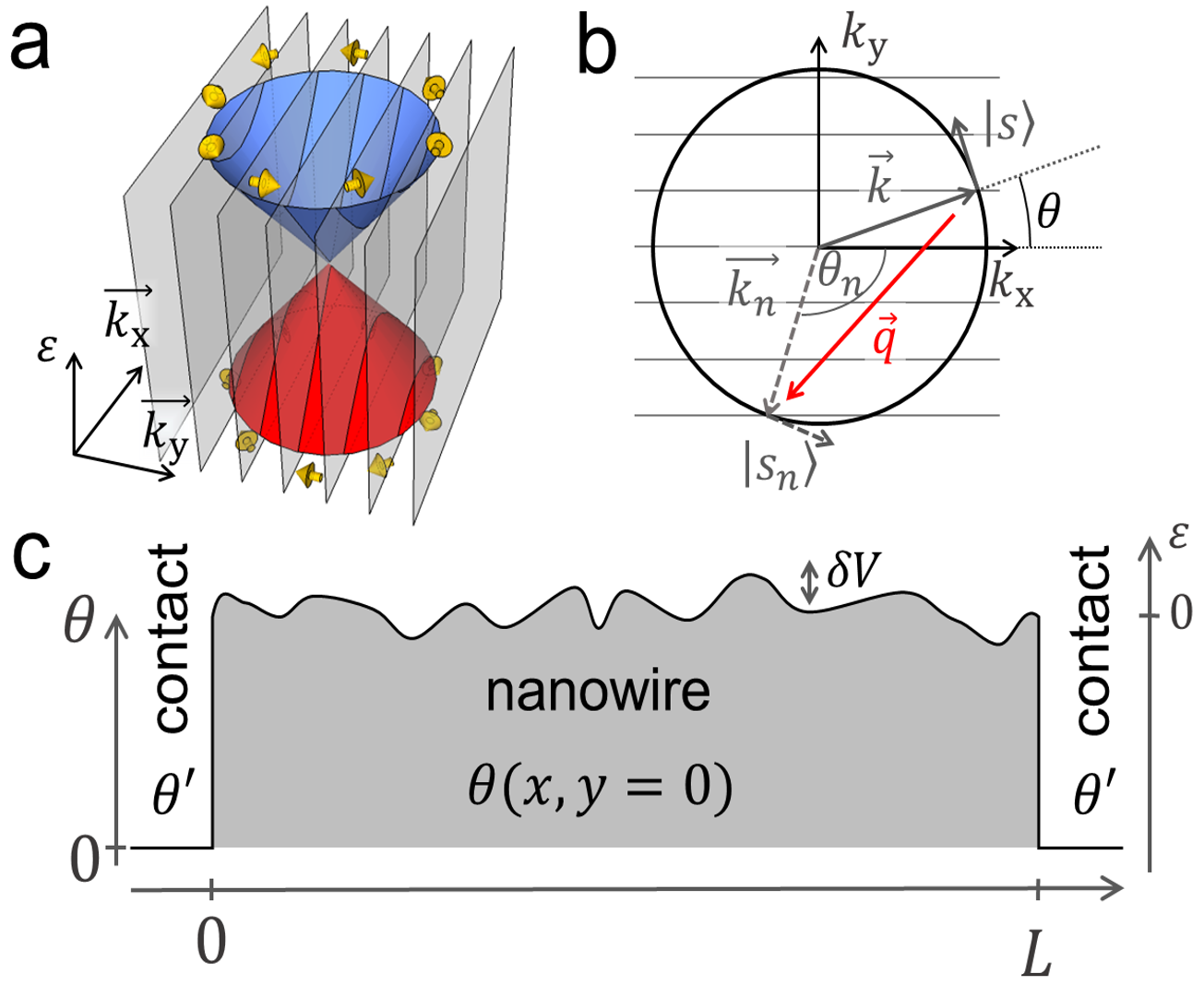}
	\caption[width=\columnwidth]{a) 2D band structure of a massless fermion system with spin-momentum locking. The planes correspond to the section of the cone for discrete values of the transverse wave vector due to quantum confinement. b) section of the 2D band structure at a given Fermi energy. c) 1D cut of the disorder and contact potential at $y=0$, plotted together with the corresponding $\theta$ angle.} 
	\label{fig1}
\end{figure}

For clarity, we consider the case of a band structure with a single spin-helical Dirac cone (Fig.~\ref{fig1}a), as realized for example in Bi$_2$Se$_3$ \cite{Xia2009}, but our results are easily generalized to the case of graphene nanoribbons or carbon nanotubes in absence of intervalley scattering. The system is described by the Dirac Hamiltonian
\begin{equation}
\mathcal{H}=v\mathbf{p}\cdot {\boldsymbol\sigma} + V(\mathbf{r}) + V_c
\label{Hamiltonian}
\end{equation}
with ${\boldsymbol\sigma}=(\sigma_x,\sigma_y)$ the Pauli sigma matrices, $\mathbf{r}=(x,y)$ where $x$ is the longitudinal coordinate and $y$ the transverse coordinate, and $V(\mathbf{r})$ stands for a Gaussian correlated scalar disorder such that
\begin{equation}
\langle V(\mathbf{r})V(\mathbf{r'})\rangle=g\frac{(\hbar v)^2}{2 \pi \xi^2}e^{-|\mathbf{r}-\mathbf{r'}|^2/2 \xi^2}.
\label{disorder}
\end{equation}
Here $\xi$ is the disorder correlation length and $g$ is a dimensionless parameter that measures the disorder strength. The qualitative results of our study do not depend strongly on the exact nature of the disorder correlation function. The contacts are modeled by a potential $V_c \rightarrow -\infty$ for $x<0$ and $x>L$ and $V_c=0$ otherwise \cite{Tworzydlo2006,Bardarson2010} (see Fig.~\ref{fig1}c). This model corresponds to a strong doping of the topological insulator below metallic electrodes, or, equivalently, to the injection of quasiparticles from a metallic electrode with $k_\text{x} \gg k_\text{y}$ ($k_\text{x}$ and $k_\text{y}$ are the component of the wave vector $\mathbf{k}$ parallel and perpendicular to the axis of the nanowire). Our results, as we will see below, can be generalized to a broader class of contacts.

In a nanowire geometry, $k_\text{y}$ is quantized due to periodic boundary conditions (Fig.~\ref{fig1}a and b). In general, $k_{\text{y}}=k_{\text{n}}= \varepsilon_\text{n}/\hbar v=(n+\phi/\phi_0-1/2)\times \Delta/\hbar v$ where the $1/2$ comes from the Berry phase induced by spin-momentum locking on a curved surface \cite{Zhang2009b,Rosenberg2010,Ostrovsky2010} (absent in graphene nanoribbons), $n \in \mathbb{Z}$ is a mode index, $\phi$ is the magnetic flux threaded through the cross section of the nanowire, $\phi_0=h/e$ is the magnetic flux quantum and $\varepsilon_\text{n}$ is the energy of the mode $n$. In addition to the transverse quantized energy $\Delta$, we also consider the longitudinal quantized energy $\Delta_\parallel=\pi \hbar v/L$ where $L$ is the length of the wire.

Ignoring first the quantum confinement and considering the 2D limit only, the transport relaxation time $\tau$ and the transport length $\ell=v\tau$ can be explicitly determined for a Gaussian potential, starting from Fermi's golden rule\cite{Adam2009,Bardarson2010}. As expected, $\ell$ and $\tau$ do not depend on the incident direction of the $\mathbf{k}$-vector of the wave function \footnote{In agreement with \cite{Adam2009}, we found a difference of a factor 4 with respect to the expression of $\ell$ written in \cite{Bardarson2010}}:
\begin{equation}
\ell=v\tau=\frac{2k\xi^2}{g}\frac{\exp(k^2\xi^2)}{I_1(k^2\xi^2)}
\label{Transport length}
\end{equation}
where $I_1$ is the modified Bessel function of the first kind. Even if the 2D limit is valid only for $k\ell \gg 4\pi$, such that the divergence of $\ell$ at low energy is smoothed out, $\ell$ reaches a minimum $\ell_\text{m}$ at an energy $\varepsilon$ corresponding to $k\xi\sim 1$. As a result, for a conductor with a finite width $W < \ell_\text{m}$, the disorder is not strong enough to set the system in the 2D diffusive limit. Boundary conditions modify the density of states that exhibits a maximum at each transverse mode opening, a feature typical of the 1D nature of the subband associated to the mode: the system is then quantum confined. This condition reads $g \lesssim 0.5$ for $\xi/W=0.05$ (see Appendix~\ref{ltrDependence}). We define the different transport regimes by comparing $\ell_\text{m}$ to the length and perimeter of the device. The ballistic, quasi-ballistic and diffusive regimes correspond to $\ell_\text{m} > 2L$, $2L > \ell_\text{m} > W$ and $W > \ell_\text{m}$ respectively. A factor 2 is introduced due to the different boundary conditions along the longitudinal and transverse directions of the nanowire. 

\section{Transmission modes and disorder}
\label{TransDisorder}

At high energies, $\ell$ is larger than the system size and transport properties are determined by the interface between the nanowire and the lead. A perfect interface simply consists of a step in the chemical potential at $x=0$ and $x=L$. This corresponds to the case of a nanowire with screening length shorter than the Fermi wavelength, a good approximation for real systems \cite{Dufouleur2016}. As long as the step does not depend on $y$, translational invariance implies the conservation of $k_\text{y}$ such that only intra-mode backscattering processes takes place at the interface. Each transverse mode can be considered independently following Refs.~\onlinecite{Katsnelson2006,Tworzydlo2006}. For a given mode $n$ with $\theta = \arctan ([(\varepsilon/\varepsilon_\text{n})^2-1]^{-1/2})$ (see Fig.~\ref{fig1}b), the reflection ($r_{\theta',\theta}$) and transmission ($t_{\theta',\theta}$) coefficients only depend on $\theta$ in the nanowire and $\theta'$ in the contact. Contrary to the massive case, the Hamiltonian (\ref{Hamiltonian}) does not require the continuity of the spatial derivative of the wave function but the continuity of the two-component wave function, which gives $r_{\theta',\theta}=\sin([\theta-\theta']/2)/\cos([\theta+\theta']/2)$ and $t_{\theta',\theta}=\cos(\theta')/\cos([\theta+\theta']/2)$. The total transmission amplitude $t_\text{n}$ of the mode $n$ considering a  non-disordered nanowire and two contacts with perfect interfaces is
\begin{equation}
t_\text{n}=\frac{\cos \theta \cos \theta'}{\cos \theta \cos \theta' \cos \varphi +i \left(\sin\theta\sin\theta'-1\right)\sin\varphi}
\label{Transmission amplitude}
\end{equation}
with $\varphi=k_\text{x} L$. This expression is a generalization to an arbitrary $\theta'$ of the transmission amplitude for propagating modes found in Ref. \citenum{Tworzydlo2006} (where $\theta'=0$). Contrary to Ref. \citenum{Tworzydlo2006}, we do not consider evanescent modes which exponentially vanish far from the Dirac point and for long distances between the contact ($L > W$) such that transport properties are strongly dominated by propagating modes. The transmission is $T_\text{n}=\lvert t_\text{n} \rvert^2$.

We first consider a weak disorder and follow a Wentzel-Kramers-Brillouin approach \cite{Shankar1994}. The main effect of the disorder is to randomly redistribute the phase $\varphi$ of each mode without inducing any inter-mode scattering. Thus, the position of the Fabry-Pérot resonances that corresponds to $k_\text{x} L=p\pi$ ($p\in \mathbb{N}$) in the clean case will be shifted depending on the disorder configuration and the disorder-averaged transmission is given by $\langle T_\text{n} \rangle=1/2\pi \times \int_0^{2\pi} \left| t_\text{n}(\varphi) \right|^2 d(\varphi)$. This approach is equivalent to a temperature smearing with $4k_\text{B}T > \Delta_\parallel$. $\langle T_\text{n} \rangle$ can be explicitly calculated:
\begin{equation}
\langle T_\text{n} \rangle=\frac{\cos \theta \cos \theta'}{1-\sin \theta \sin \theta'}.
\label{Mean_Trans_theta_c}
\end{equation} 
More particularly, when $V_c \rightarrow -\infty$ we have $\theta'\simeq 0$ and
\begin{equation}
\langle T_\text{n} \rangle=\cos \theta=\sqrt{1-\left(\frac{\varepsilon_\text{n}}{\varepsilon}\right)^2}.
\label{Mean_Trans}
\end{equation} 
We note that the transmission of a mode differs from 1 except for $\theta=0$ that corresponds to the perfectly transmitted mode discussed in Refs. \citenum{Ando:2002es}, \citenum{Bardarson2008a} and \citenum{Bardarson2010}. This mode corresponds to $\varepsilon_\text{n}=0$ which requires half a quantum of flux to be threaded through the cross-section of the nanowire to compensate the Berry phase picked up by a particle when it goes around the nanowire. This approach is valid as long as (i) the system is quantum confined, which requires $\ell > W$ and (ii) $\lvert d \lambda/dx \lvert \ll 2 \pi$ where $\lambda=2\pi/k_\text{x}$ \cite{Shankar1994}. It is therefore not valid close to the onset of a mode but it is satisfied for $\varepsilon \gtrsim \varepsilon_\text{n}$ for the conditions we are using here.

We compare the analytical expression (\ref{Mean_Trans}) to numerical simulations following the method presented in Refs. \citenum{Bardarson2007}, \citenum{Bardarson2010} and \citenum{Xypakis2017}. The transmissions of a disordered nanowire with $W=200$ nm and $L=500$ nm is calculated up to $\varepsilon/\Delta \sim 25$ and averaged over $\sim 10^3$ disorder configurations for different strengths of disorder ranging from the ballistic limit ($g=0.02$, $\ell_\text{m} > 2L$) to the diffusive limit ($g=1$, $\ell_\text{m} < W$). We choose a correlation length $\xi = 10$ nm consistent with experimental measurements \cite{Dufouleur2016}.

\begin{figure}[t]
	\centering
	\includegraphics[width=\columnwidth]{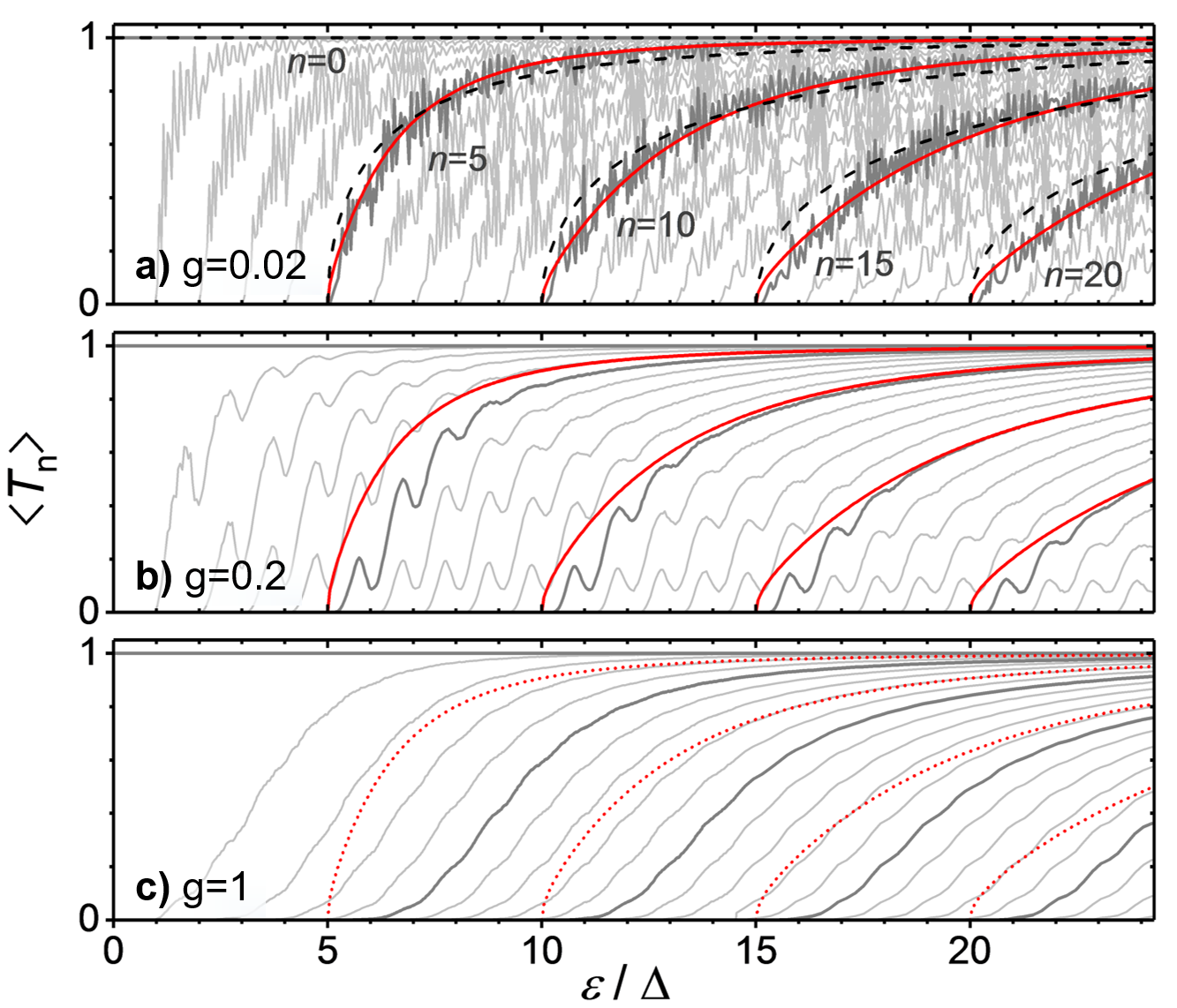}
	\caption[width=\columnwidth]{The transmission of the transverse modes calculated and disorder-averaged for different disorder strength $g$ (0.02, 0.2 and 1) ranging from the ballistic to the diffusive regime. Black dashed lines indicate the ballistic transmission with no quantum corrections ($\gamma_0=0$ and $\gamma_1=0$). Red lines are the best fit of the transmission including quantum corrections for a) ($\gamma_0=0.39$ and $\gamma_1=0.43$). Good agreement with numerical data is obtained in b) with $\gamma_0=0.4$ and $\gamma_1=-0.85$ for $g=0.2$, whereas no satisfactory parameters fit the $g=1$ data; the $\gamma_0=0.4$ and $\gamma_1=-0.85$ transmissions are indicated by red dotted lines in c).}
	\label{fig2}
\end{figure}

Results are presented in Fig. \ref{fig2} for half a flux quantum threading the cross section of the nanowire. Due to time reversal symmetry, the zero energy mode is topologically protected and its transmission is equal to 1 independently of the disorder strength\cite{Bardarson2010,Bardarson2008a}. All other modes exhibit Fabry-Pérot resonances that are not fully averaged out for a weak disorder ($g \lesssim 0.02$). Nevertheless, the average transmission roughly follows the analytical formula (\ref{Mean_Trans}) in general. 

We notice two kinds of deviation from the adiabatic limit for the simulated data. Firstly, dips appear close to the onset of each mode for $g \gtrsim 0.05$ (see for instance Fig.~\ref{fig2}b)). They are smeared out for $g \gtrsim 1$. This is the signature of the modification of the density of states by quantum confinement.  More quantitatively, starting from the Fermi golden rule and assuming  $\ell > W$, we obtain:
\begin{equation}
\frac{1}{\tau_\textbf{k}}=\frac{g v}{2W} \sum_{\textbf{k}_\text{n}} \left|\frac{1-\cos^2\theta_\text{q}}{\cos \theta_\text{n}}\right| \exp \left(-\frac{q^2\xi^2}{2}\right)
\label{tau Quasiballistic}
\end{equation}
where $\tau_\textbf{k}$ is the transport time of a $\textbf{k}$-state, $\textbf{q}=\textbf{k}_\text{n}-\textbf{k}$, $\theta_\text{n}$ is the angle of final $\textbf{k}_\text{n}$-state and $\theta_\text{q}=\theta-\theta_\text{n}$. Contrary to the diffusive case, $\tau_\textbf{k}$ explicitly depends on the initial state $\textbf{k}$. Dips in the transmission come from the opening of transverse mode associated to the divergence of the 1D density of state ($\cos \theta_\text{n}=0$). Far from the onset $(\varepsilon-\varepsilon_n \gg \Delta)$, this effect is strongly reduced by the exponential cut-off of the Gaussian disorder ($q^2 \xi^2 /2 \gg 1$). For $g \gtrsim 1$, the disorder broadening induces overlapping dips and the transmission deviates clearly from its ballistic limit (see Fig. \ref{fig2}c). This corresponds to $\delta V = \langle V(\mathbf{r})V(\mathbf{r})\rangle^{1/2} \sim \Delta$.

Secondly, the transmissions $T_\text{n}$ is not strongly affected by a weak disorder and $T_\text{n}$ remains close to its clean limit $T_\text{n}^\ast$ ($g=0$) given by Eq.~\eqref{Mean_Trans}, for $g$ up to $0.2$. Nevertheless, some deviations $\delta T_\text{n}=T_\text{n}-T_\text{n}^\ast$ due to the coupling between the different transverse modes are measured for $g\neq 0$. Despite this coupling and for a weak disorder ($g \lesssim 0.2$), $\delta T_\text{n}$ is small and the transport properties can still be described by the transmission of the transverse modes $\textbf{k}_\text{n}$. In this case, we observe two kind of systematic deviations:
\begin{enumerate}
	\item close to the onset ($\varepsilon \gtrsim \varepsilon_\text{n}$), $\delta T_\text{n} < 0$. Indeed, for a Gaussian disorder, the scattering is limited to adjacent modes due to the exponential cut-off in Eq.~\ref{tau Quasiballistic}. It will be dominated by adjacent mode close to the onset since the density of states diverge at the onset (the $\cos \theta_\text{n}$ term in Eq.~\ref{tau Quasiballistic}). Therefore, scattering is strong when $\varepsilon \gtrsim \varepsilon_\text{n}$, which generally reduces the transmission.
	\item at high energy ($\varepsilon \gg \varepsilon_\text{n}$), $\delta T_\text{n} > 0$. In this regime, the $1-\cos^2 \theta_\text{q}$ term in Eq.~\ref{tau Quasiballistic} vanishes: single scattering events should have no impact on the transmission and should lead to  $\delta T_\text{n} \rightarrow 0$. This deviation can be only explained by quantum interferences (multiple scattering processes) inducing conductance corrections. Simulations give disorder averaged transmissions which average out any quantum interferences contributions except for weak anti-localization (WAL) interferences that lead to some positive corrections \cite{Ando1998}.
\end{enumerate}

\begin{figure}[t]
	\centering
	\includegraphics[width=\columnwidth]{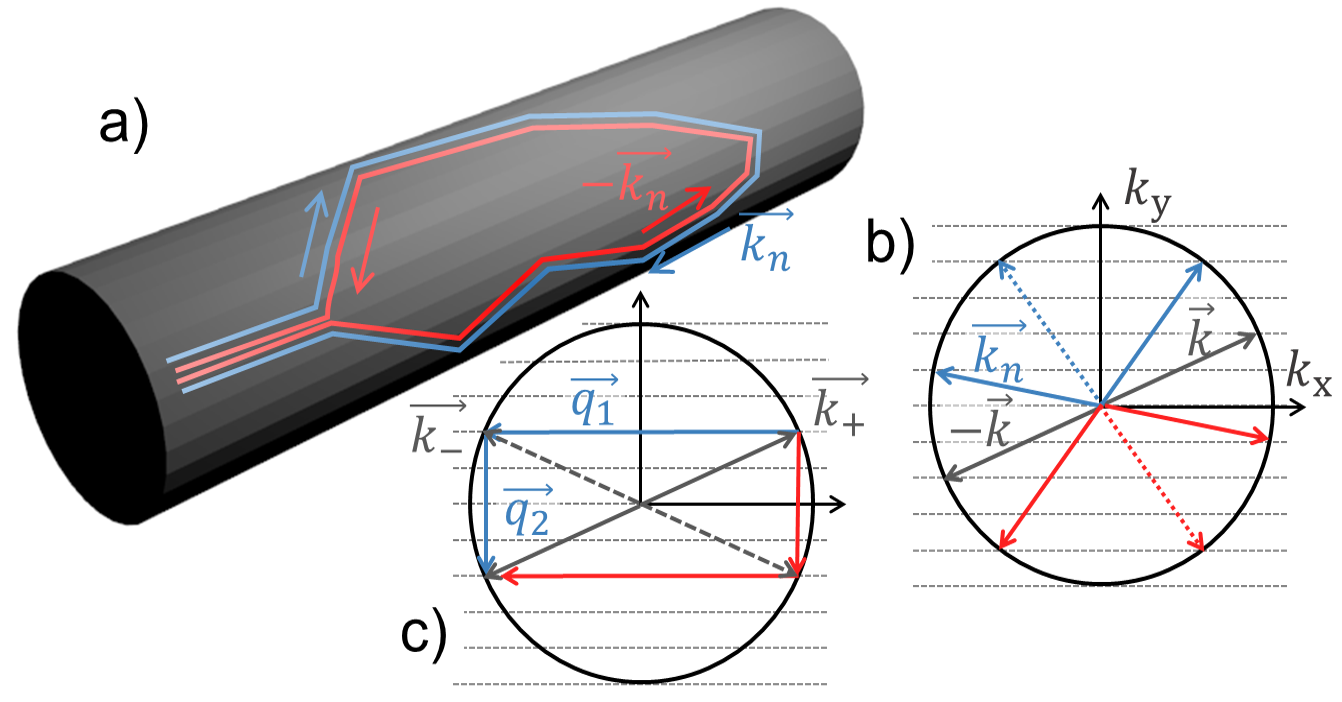}
	\caption[width=\columnwidth]{a) Schematic of a loop corresponding to WAL quantum interferences in the real space. b) The same loop represented in the $\textbf{k}$-space. c) The loops that dominate the WAL corrections for a mode $n$ (we chose here $n=2$ and $\phi=\phi_0/2$)}
	\label{fig3}
\end{figure}

We can roughly estimate the energy dependence of the corrections due quantum interferences. The trajectories that are generally involved in WAL loops are $\textbf{k} \rightarrow \textbf{k}_1 \rightarrow \cdots \rightarrow \textbf{k}_\text{n} \rightarrow -\textbf{k}$ and their time reversed trajectories $\textbf{k} \rightarrow -\textbf{k}_\text{n} \rightarrow \cdots \rightarrow -\textbf{k}_1 \rightarrow -\textbf{k}$ as described in the figure \ref{fig3}. On the one hand, the exponential cut-off in Eq.~(\ref{tau Quasiballistic}) favors loops minimizing the number of inter-mode scattering events in the WAL corrections and the minimal exponential cut-off is obtained for only two scattering events $\textbf{q}_1$ and $\textbf{q}_2$. In this case, we have $\exp \left(-(q_1^2+q_2^2)\xi^2/2\right)=\exp \left(-q^2\xi^2/2\right)=\exp \left(-2 {k}^2\xi^2 \right)$. The amplitude of probability of such a loop is even strongly enhanced if one of the two scattering events implies a reflection on the interface between a contact and the nanowire (Fig.~\ref{fig3}c)). Indeed, the reflection coefficient $r$ of the interface for $\theta'=0$ is given by $r_\text{n}=\sin (\theta/2)/\cos (\theta/2)=\sin(\theta)/(1+\cos (\theta))$, which tends to zero when $\varepsilon \gg \varepsilon_\text{n}$ as a power law of $\varepsilon$ instead of an exponential decay $\left(\propto \exp \left(-2 \varepsilon^2\xi^2/\hbar^2 v^2 \right) \right)$ for a disorder-driven scattering process. On the other hand, the spin projection of the initial state on the scattered state tends to favor the contribution of loops that maximize the $1-\cos^2 \theta_\text{q}$ term in Eq.~(\ref{tau Quasiballistic}). This is obtained for loops involving again two scattering events only ($\textbf{q}_1$ and $\textbf{q}_2$) with $\theta_{\text{q}_1}=\theta_{\text{q}_2}=\pi/2$. Nevertheless, the contribution of such trajectories to quantum corrections is exponentially suppressed and the WAL corrections are dominated at high energy by loops implying one reflection on the contact interface and one disorder-driven scattering event as shown in Fig.~\ref{fig3}c). 

The quantum interferences involved in WAL are then dominated by trajectories, which amplitude of probability is given by the product of $r_\text{n}$ (related to $\textbf{q}_1$), and the amplitude of probability associated to the $\textbf{q}_2$ inter-mode scattering event (see Fig.~\ref{fig3}c)). The latter is related to the scattering probability given by Eq.~(\ref{tau Quasiballistic}) with $q_2=2n\pi/W$ that is energy independent and is therefore an even function of $\varepsilon_\text{n}/\varepsilon$. The energy dependence of $r_\text{n}$ reads
\begin{equation}
r_\text{n}=\frac{\sin \left( \theta \right)}{1+\cos \left( \theta \right)}=\frac{\varepsilon_\text{n}/\varepsilon}{1+\sqrt{1- \left(\varepsilon_\text{n}/\varepsilon \right)^2}},
\label{rn}
\end{equation}
an odd function of $\varepsilon_\text{n}/\varepsilon$. Hence, the amplitude of probability of the trajectories is an odd function of $\varepsilon_\text{n}/\varepsilon$ and the probability an even function of $\varepsilon_\text{n}/\varepsilon$ that can be expanded at high energy ($\varepsilon_\text{n}/\varepsilon \ll 1$) in a Taylor series. Due to the parity, the WAL quantum corrections to the transmission only involve even order of $\varepsilon_\text{n}/\varepsilon$ and the transmission can be generally approximated at high energy by
\begin{equation}
\langle T_\text{n} \rangle=\sqrt{1-\left(\frac{\varepsilon_\text{n}}{\varepsilon}\right)^2} \times \left(1+\sum_{p=0}^{\infty}\gamma_p \left(\frac{\varepsilon_\text{n}}{\varepsilon}\right)^{2(p+1)}\right)
\label{Trans plus Corr}
\end{equation} 
where the coefficients $\gamma_p$ depend on the effectiveness of the disorder in coupling $\mathbf{k}_+$ to $-\mathbf{k}_-$ (see Fig.~\ref{fig3}). When we consider only the large energy limit corresponding to $\varepsilon \gg \varepsilon_\text{n}$, only the lowest order in $(\varepsilon_\text{n}/\varepsilon)^2$ will significantly contribute to the quantum corrections.
We find good agreement with the numerical simulation by including only the first two order corrections $\gamma_0$ and $\gamma_1$ for the full set of transmissions up to $g=0.5$ at high energy (see Fig.~\ref{fig2}).

Importantly, for a weak disorder, the transmissions $\langle T_\text{n} \rangle$ only depend on the mode index $n$ through the ratio $\varepsilon_\text{n} / \varepsilon$ such that the $\gamma$ parameters do not depend on $n$ anymore as shown in the Appendix~\ref{GammaParameters}. This allows us to describe all the transmissions with a single set of $\gamma$ parameters. Also, $\gamma_0$ and $\gamma_1$ only weakly depend on $L$, $\xi$ and $g$ as long as $\ell>W$ (see Appendix~\ref{GammaParameters}) with $\gamma_0 \approx 0.43$ and $\gamma_1 \approx -0.93$. Hence, transport properties of the system depend neither on the length nor on the detail of the microscopic disorder ($g$ and $\xi$) anymore.

\section{Conductance and shot noise}

From the transmissions of the different transverse modes, we extract the conductance $G=e^2/h \times \sum_\text{i} \langle T_\text{i} \rangle$ and the Fano factor $F=\sum_\text{i} \langle T_\text{i} \rangle (1-\langle T_\text{i} \rangle)/ \sum_\text{i} \langle T_\text{i} \rangle$ for a quantum confined nanowire. For a large number of modes ($\varepsilon \gg \Delta$), we can take the continuous limit and we replace the discrete sum by an integral where $i$ is considered as an continuous index. Keeping only $\gamma_0$ and $\gamma_1$ corrections, we have
\begin{gather}
G=\frac{e^2}{h}\frac{2\varepsilon}{\Delta}\mathcal{T},
\label{Conductance}\\
F=1-\left(\frac{2}{3}+\frac{4}{15}\gamma_0+\frac{4}{35}\gamma_1\right)/\mathcal{T}
\label{FanoFactor}
\end{gather}
with the average transmission per transverse mode
\begin{equation}
\mathcal{T}=\frac{\pi}{4}+\frac{\pi}{32}\left(2\gamma_0+\gamma_1\right).
\label{MeanTrans}
\end{equation} 
Agreement with the simulations is excellent for $g<0.05$ even at low energy and it remains very good up to $g \simeq 0.5$ in the limit of large number of modes. We note that for all $g$, best fits gives $2\gamma_0 \sim -\gamma_1$ such that the conductance and the Fano factor can be approximated within an error $<1\%$ by
\begin{align}
G&\simeq\frac{e^2}{h}\frac{2\varepsilon}{\Delta}\frac{\pi}{4}
\label{ConductanceApprox}\\
F&\simeq 1-\frac{8}{3\pi}
\label{FanoFactorApprox}
\end{align}

\begin{figure}[t]
	\centering
	\includegraphics[width=\columnwidth]{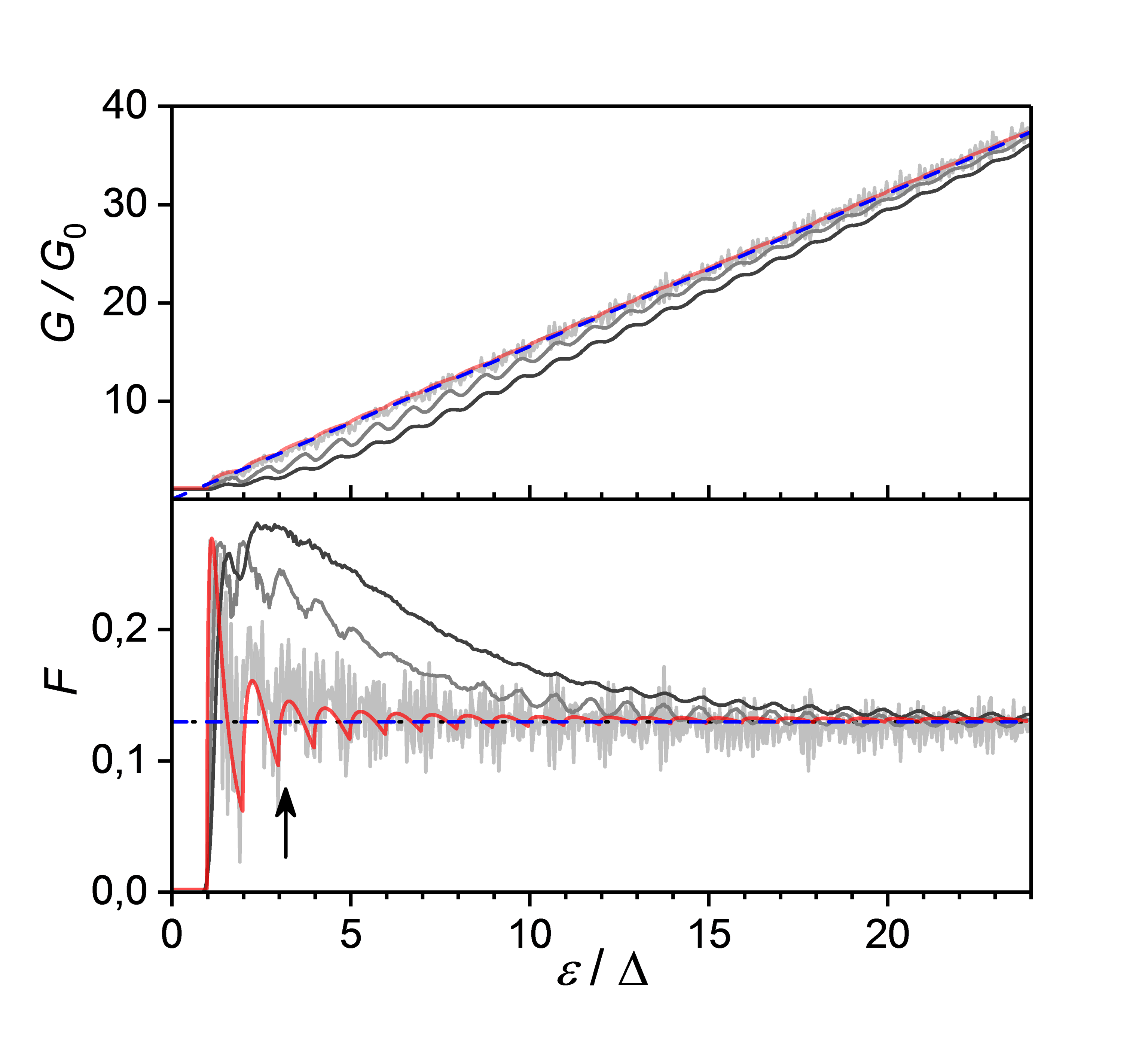}
	\caption[width=\columnwidth]{Upper panel: Energy dependence of the conductance. Gray lines are numerical data for $g=0.02$, $0.2$ and $0.5$ (from light to dark gray). The red line is the conductance calculated from Eq.~\eqref{Conductance} with $\gamma_0=0.43$ and $\gamma_1=-0.92$ and blue dashed line is the large number of mode limit for the same values of $\gamma_0$ and $\gamma_1$. Lower panel: Energy dependence of the Fano factor shown with the same color code as for the conductance. The ballistic limit ($\gamma_0=\gamma_1=0$) is indicated by the dotted line and the arrow points at the energy corresponding to $k\xi=1$.}
	\label{fig4}
\end{figure}

As already reported for graphene nanoribbons \cite{Tombros2011,Terres2016}, the linear dependence of the conductance with $\varepsilon$ or $k_\text{F}$ is the signature of quantum confinement and for the model we used for the contact, the proportional factor is $\mathcal{T}=\pi/4$. In the diffusive limit indeed, the conductivity $\sigma$ is related to the density of states $\partial n / \partial \mu$ and to $\ell$ through the Einstein relation: $\sigma=e^2 (\partial n/\partial \mu) v \ell /2$. At low energy, the disorder is short-range and $k\xi \ll 1$. In this regime, $\ell \propto \varepsilon^{-1}$ and does not depend on $\xi$ anymore (see Appendix~\ref{ltrDependence}). As $\partial n/\partial \mu \propto \varepsilon$, the conductance does not show any energy dependence. At high energy, the disorder is long-range ($k\xi \gg 1$) and $\sigma \propto (\varepsilon\xi)^3$. In our model, it is not possible to know the exact energy dependence of the conductance since $\xi$ is a given parameter. Nevertheless, a microscopic model \cite{Hwang2008,Culcer2010} considering an ensemble of charge impurities with screened Coulomb potential (a non-Gaussian potential) gives $\sigma \propto \varepsilon^2$ in agreement with experimental results.

\begin{figure}[t]
	\centering
	\includegraphics[width=\columnwidth]{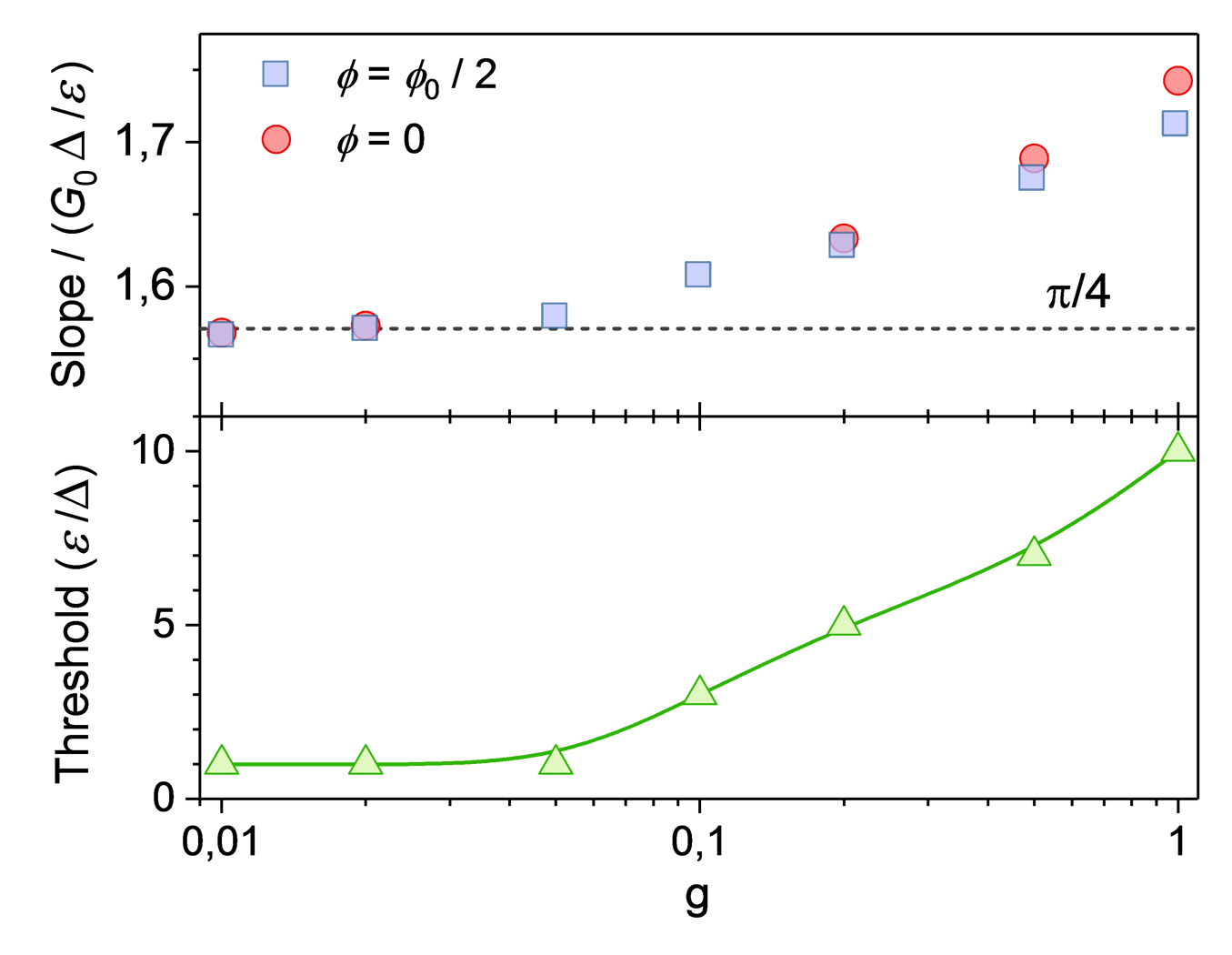}
	\caption[width=\columnwidth]{Upper panel: slope of $G(\varepsilon/\Delta)$ obtained by directly fitting the conductance in the linear regime. The dashed line indicate the clear limit corresponding to $\mathcal{T}=\pi/4$ or, equivalently, to a slope of $\pi/2$. Lower panel: value of the energy from which the $G(\varepsilon/\Delta)$ is linear.}
	\label{fig5}
\end{figure}

In the simulations, deviations to the linearity are observed for $g\gtrsim 0.1$ and are maximal when $k\xi \sim 1$. Increasing the energy such that $k\xi \gg 1$ (long-range disorder), $G(\varepsilon)$ tends to its non-disordered limit (see Fig.~\ref{fig4} and \ref{fig6}). The linearity is recovered even for $g=1$ at $\varepsilon > 10 \times \Delta $, with a slope that corresponds to a $\mathcal{T}$ slightly above $\pi /4$. This is a consequence of the energy dependence of $\ell$ that reaches a minimum when $k\xi \sim 1$ and continuously increases with the energy for $k\xi \gtrsim 1$ (see Fig.\ref{SI-fig1} in Appendix~\ref{ltrDependence}). For $g=0.5$, we found $\ell > W$ for $\varepsilon/\Delta > 4$ and $g=1$, $\ell > W$ for $\varepsilon/\Delta > 6$, in rough agreement with the threshold measured in Fig.~\ref{fig5} (bottom). This goes with a slight enhancement of the slope as shown in Fig.~\ref{fig5} (top). For $g \gtrsim 0.5$, the non-linearity becomes significant and corresponds to the condition $\ell_\text{m} \lesssim W$, which determines the diffusive limit.

Generally, the transmission of a mode $\langle T_\text{n} \rangle$ depends on the model used for the contact far from the Dirac point\cite{Robinson2007,Blanter2007,Pieper2013}. Nevertheless, as long as the transmission ($t_\text{n}$) and reflection ($r_\text{n}$) coefficients of the contact-to-nanowire junction are functions of the energy through the ratio $\varepsilon/\varepsilon_n$ only, the high energy conductance and the Fano factor can be expressed in a similar way to that in Eq. (\ref{ConductanceApprox}) and Eq. (\ref{FanoFactorApprox}). Indeed, the transmission $\langle T_\text{n} \rangle$ is then also a function of $\varepsilon/\varepsilon_n$ and if we note 
\begin{equation}
	\langle T_\text{n} \rangle=f(\varepsilon_n / \varepsilon)
	\label{Condition}
\end{equation}
we have
\begin{align}
G&\simeq\frac{e^2}{h}\frac{2\varepsilon}{\Delta}\int_0^1 f(x)dx,
\label{ConductanceGeneral}\\
F&\simeq 1-\frac{\int_0^1 \left(f(x)\right)^2 dx}{\int_0^1 f(x)dx}.
\label{FanoFactorGeneral}
\end{align}
The average transmission is given by $\mathcal{T}=\int_0^1 f(x)dx$. In the model we used so far, this condition is satisfied as long as we neglect intermode scattering.  instead of 1 when the contact are macroscopic pads adiabatically coupled to the nanowire.

We can now generalize the conditions related to the energy dependence of $t_\text{n}$ and $r_\text{n}$ to any value of $V_\text{c}$ and examine the $\varepsilon$-dependence of $t_\text{n}$ and $r_\text{n}$ when the Fermi energy is constant in the contact. This corresponds to a gated device for which the field effect is screened under the contact and not in the nanowire. We consider again the case of an interface inducing no intermode scattering ($\partial V_\text{c}/\partial y=0$). This case is equivalent so set the value of $\theta'$ to a constant and to study the dependence of the transmission $T_\text{n}$ with $\theta$ as in the section \ref{TransDisorder}. In this case, the reflexion ($r_\text{n}$) and transmission ($t_\text{n}$) coefficients only depend on $\theta$ and $\theta'$. The global transmission $T_\text{n}$ depends on the energy only through $\theta$ and is therefore a function of $\varepsilon_\text{n}/\varepsilon$ only so that the condition given by Eq.~(\ref{Condition}) is satisfied.

Remarkably, in addition to the linearity of $G(\varepsilon)$, the length and disorer strength dependence of the transport properties also strongly deviate from the diffusive limit, for which $G \propto g^{-1}$ and $G \propto L^{-1}$ (see Appendix~\ref{LengthDependence}) for $g$ up to $0.5$. For $\ell > W$, scattering to modes close to their onset dominates (Eq. (\ref{tau Quasiballistic})), inducing oscillations of the transmission. As long as the disorder broadening $\delta=h/\tau$ is smaller than $\Delta$, scattering has only a marginal effect on $G$ since the non-overlapping condition $\Delta/\delta=\ell/W>1$ holds for any mode index and therefore over the full energy scale. Hence, quantum confinement drives the system in the ballistic regime. This weakening of the scattering by the quantum confinement leads to a conductance that is not proportional to the length between the contact as observed in Ref.~\citenum{Hong2014}. The ballistic feature of the conductance is confirmed by the low value of $F$ that is significantly smaller than its diffusive value $F=1/3$ at large energy (see Fig.~\ref{fig3}). This value is given by the nature of the interface between the nanowire and the contact. In the case of a interface transmission coefficient $t_\text{n}=1$ when $\varepsilon > \varepsilon_\text{n}$, like for an adiabatic quantum point contact (we have then $V_\text{c}=0$) \cite{Lin2008,Tombros2011,Terres2016}, the Fano factor vanishes. For the perfect interface considered here ($V_\text{c} \neq 0$), the transmission of each mode is not equal to 1 but $F$ is nevertheless considerably reduced with respect to its diffusive limit.

The consequences of the quantum confinement on the transport regime are specific to Dirac systems. For massive particles with mass $m$, the quantum confinement condition reads $\ell/W \simeq \Delta_0 N/\delta>1$ where $N$ is the number of transverse modes and $\Delta_0=\pi^2\hbar^2/(mW^2)$. As the energy spacing between the modes $n$ and $n+1$ is $\simeq n\Delta_0$ for $n \gg 1$, the quantum confinement condition at large energy ($N \gg 1$) does not guarantee the non-overlapping between two consecutive transverse modes, especially for small index modes ($n \simeq 1$). As a result strong deviation from the ballistic regime is expected for massive particles even for $\ell \gtrsim W$. \footnote{the differences between massless and massive particles is emphasized by the energy dependence of $\ell(\varepsilon)$. Indeed, $\ell>\ell_\text{m}$ for massless particles and quantum confinement can be ensured for the full energy scale. On the contrary, for massive particles $\ell \rightarrow 0$ when $\varepsilon \rightarrow 0$ and overlapping cannot be avoided in the low energy limit.}

\section{Aharonov-Bohm oscillations}

The influence of an Aharonov-Bohm flux on the transmission of the mode $n$ is entirely contained in the value of $\varepsilon_\text{n}=(n+\phi/\phi_0-1/2)\times \Delta$. We neglect here the effect of a Zeeman coupling that only shifts the position of the Dirac point and renormalizes the value of the Aharonov-Bohm period.

\begin{figure}[t]
	\centering
	\includegraphics[width=\columnwidth]{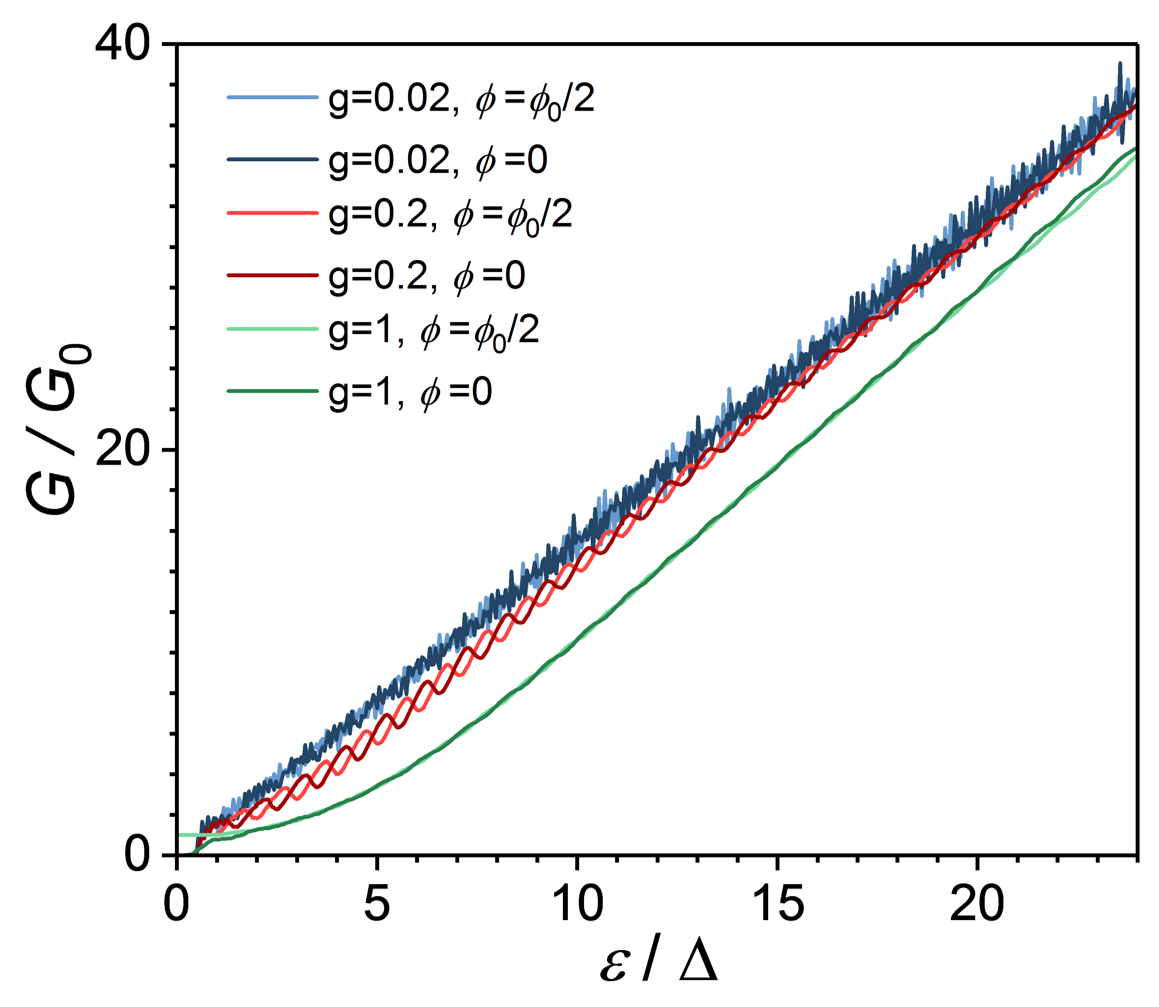}
	\caption[width=\columnwidth]{Energy dependence of the conductance for different disorder strength ($g=0.02$, $0.2$ and $1$) and for $\phi=0$ (dark colors) and $\phi=\phi_0/2$ (light colors).}
	\label{fig6}
\end{figure}

Apart from Aharonov-Bohm oscillations, the transport properties should generally not depend on the magnetic flux $\phi$ in the high energy limit, for $\varepsilon \gg \Delta$. For $\varepsilon \lesssim 2\Delta$, the transport properties are strongly influenced by the existence of a perfectly transmitted mode\cite{Bardarson2010} ($\varepsilon_\text{n}=0$ for $n=0$ and $\phi=\phi_0/2$), which has a topological origin and is therefore insensitive to the disorder strength $g$, the length $L$ of the nanowire or any other parameter that do not break the symmetries protecting the topological class. As a result, the impact of this mode will be particularly important at low energy for strongly disordered nanowires as seen in Fig.~\ref{fig6}. For $\varepsilon > 2 \Delta$, no significant difference in the trends of the conductances can be observed for $\phi=0$ and $\phi=\phi_0/2$ and the influence of the perfectly transmitted mode on transport properties vanishes.

We notice a $\pi$-phase shift of the oscillations of $G(\varepsilon)$ when the disorder has an intermediate strength ($g=0.2$ in Fig.~\ref{fig6}). This feature is also reported in a recent work, using a different code for the simulations\cite{Ziegler2017}. In this case, disorder is strong enough to couple the transverse modes (inter-mode scattering), which generates regular dips in the transmission as already mentioned in the section \ref{TransDisorder} (see Fig.\ref{fig2}b)) but it is weak enough to avoid overlapping between the dips (Fig.~\ref{fig2}c)), which maximizes the oscillations in the conductance. As this effect is related to the opening of new transverse modes $\varepsilon_\text{n}=(n+\phi/\phi_0-1/2)\times \Delta$, the oscillations are phase-shifted for $\phi=0$ and $\phi=\phi_0$. 

We use our model to extract the complete flux dependence of the conductance at low and large energy in the clean limit where intermode scattering is neglected ($g < 0.1$). We define  $\delta G (\varepsilon)= G(\varepsilon, \phi_\text{max})-G(\varepsilon, \phi_\text{min})$ where $\phi_\text{max}$ (respectively $\phi_\text{min}$) refers to the flux that maximizes (respectively minimizes) the conductance at a given energy $\varepsilon$. As we can see in Fig. \ref{fig4}, the Aharonov-Bohm amplitude $\delta G (\varepsilon)$ is maximum at $\varepsilon=0$ where $\delta G=e^2/h$. Close to $\varepsilon=\Delta/2$, $\delta G$ drops down and oscillates in a sawtooth manner with a period $\Delta /2$ as experimentally observed in \cite{Hong2014,Jauregui2016}. More information on the specific shape of the Aharonov-Bohm oscillations are given in Appendix~\ref{app:AB}. Fig.~\ref{fig4} indicates that each period is associated to a $\pi$-phase shift which has been experimentally observed in Ref.~\citenum{Jauregui2016}. Such phase shifts are a consequence of the quantum confinement and has a trivial origin. Only the low energy pinning of $\delta G$ at $e^2/h$ for any values of $g$, $L$, $W$ or $\xi$ is a signature of the nontrivial topology.
 
\begin{figure}[t]
	\centering
	\includegraphics[width=\columnwidth]{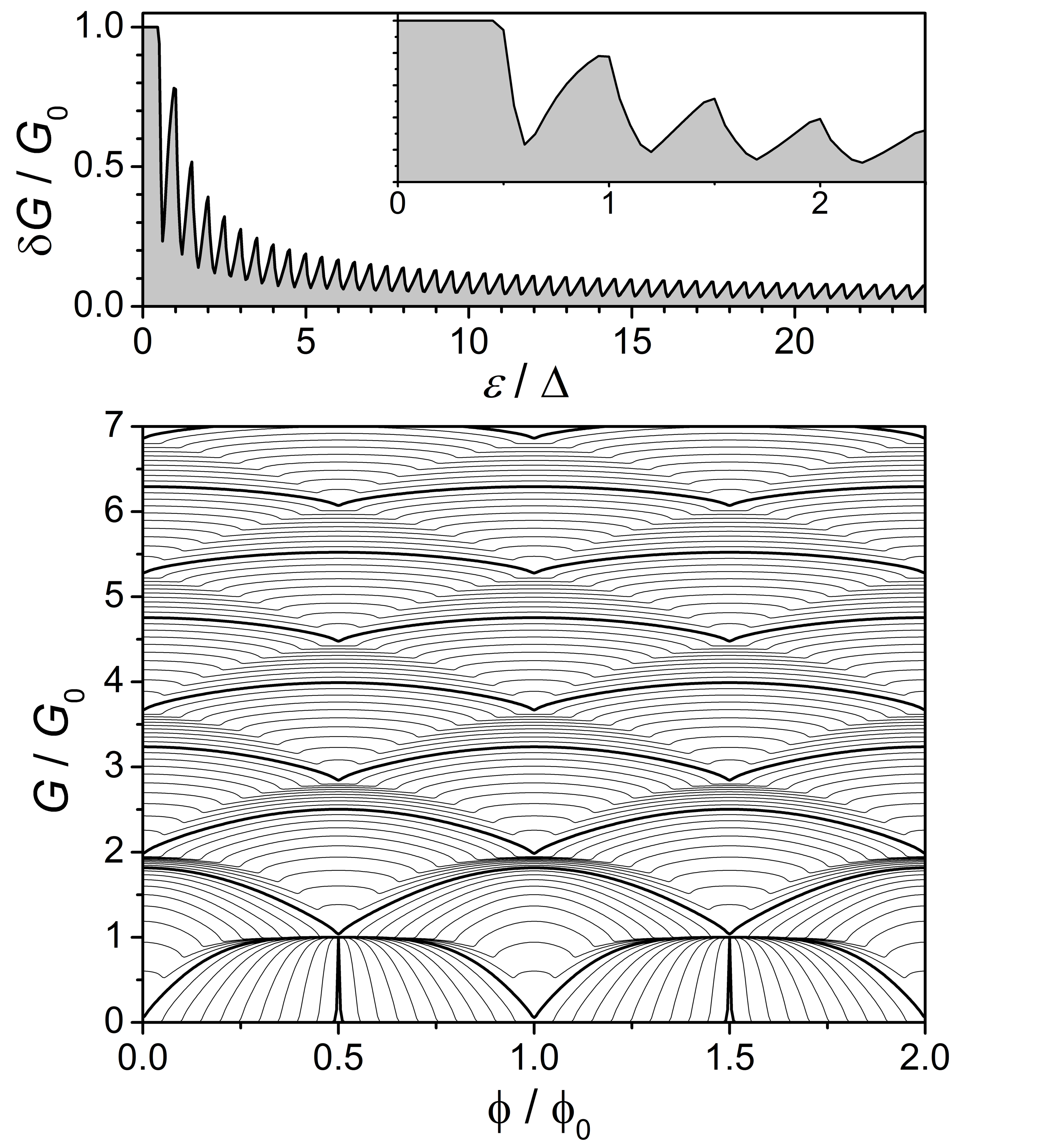}
	\caption[width=\columnwidth]{Upper panel: Energy dependence of Aharonov-Bohm oscillations for $g=0.02$ ($\gamma_0=0.43$ and $\gamma_1=-0.92$) at a temperature $4k_\text{B}T/\Delta=0.01$. The inset is a zoom closed to $\varepsilon/\Delta=0$. Lower panel: the flux dependence of the conductance for different energy at the same temperature. The smallest value of the conductance $G(\phi)$ corresponds to $\varepsilon/\Delta=0$ and the largest one to $\varepsilon/\Delta=4.5$ in steps of $\varepsilon/\Delta=0.05$. Energies such that $2\varepsilon/\Delta \in \mathbb{N}$ are indicated with bold lines.}
	\label{fig7}
\end{figure}

\section{Conclusion}

In summary, we determined the transmission of any mode of a quantum confined Dirac nanowire including quantum correction in the presence of disorder and for a perfect interface with the contact. Our analytical analysis is in good agreement with numerical simulations and shows that quantum confinement ($\ell/W>1$) drives the system into a ballistic regime, with an average transmission per mode of $\pi/4$ and a Fano factor $F\simeq 0.13$. Aharonov-Bohm oscillations are found to be periodically modulated in energy with a period corresponding to $\Delta/2$. A phase shift of the oscillations occurs every time that the Fermi energy crosses an integer value of $\Delta/2$. At low energy ($\varepsilon < \Delta/2$), a signature of the topology can be seen in the amplitude of oscillations which saturates at $e^2/h$, independently of the geometry or the microscopic properties of the disorder.
 
\subsection*{Acknowledgments}
J.D. gratefully acknowledges the support of the German Research Foundation DFG through the SPP 1666 Topological Insulators program. The DFG also supported this work through the Collaborative Research Center SFB 1143.
This work was partially supported by the ERC starting grant QUANTMATT n°= 679722.


\appendix
\section{Energy dependence of $\ell / W$}
\label{ltrDependence}

\begin{figure}[t]
	\centering
	\includegraphics[width=\columnwidth]{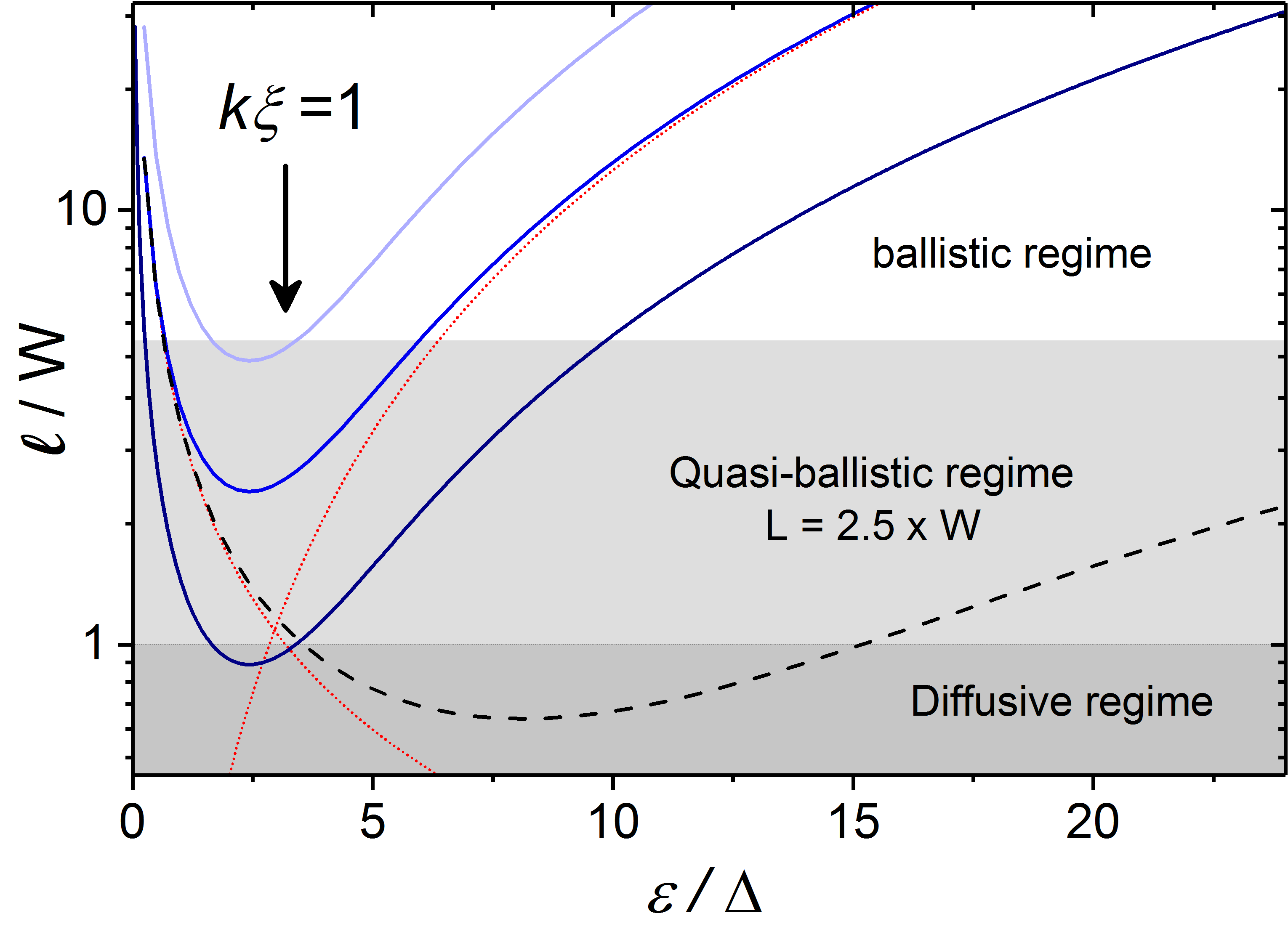}
	\caption[width=\columnwidth]{Transport length $\ell$ as a function of energy $\varepsilon$ for $\xi/W=0.05$ and $g=0.1$ (light blue), $0.2$ (blue) and $0.5$ (dark blue). Red dotted lines indicates the asymptotic dependence at low and high energies for $\xi/W=0.05$ and $g=0.2$. The black dashed line shows $\ell$ with $\xi/W=0.015$ and $g=0.2$. The gray domains indicate the ballistic regime characterized by $\ell>2L$ with $L=2.5 \times W$ and the quantum confined (quasi-ballistic) regime for which $\ell>W$.}
	\label{SI-fig1}
\end{figure}

We present in Fig.~\ref{SI-fig1} the energy dependence of $\ell/W$. For Dirac fermions, the transport length $\ell$ has a minimum $\ell_\text{m}$ for $k\xi \sim 1$ and diverges both at low energy (due to the reduction of the density of states) and at large energy (due to the anisotropy of scattering). In those two limits, $\ell$ can be approximated by its asymptotic form
\begin{equation}
\tag{A1}
\ell \sim
\begin{dcases}
4/(gk) &\text{ for } k\xi < 1,\\
2\sqrt{2\pi}k^2\xi^3/g &\text{ for } k\xi > 1.
\label{l_asmptot}
\end{dcases}
\end{equation}
This is different from the massive case for which $\ell \rightarrow 0$ at low energy. As the 2D approximation is only valid for $k\ell \gg 1$, the low energy divergence will be smoothed out.

In Fig.~\ref{SI-fig1}, the quantum confinement condition $\ell_\text{m} > W$ is satisfied for $g \lesssim 0.5$ for $\xi/W=0.05$. We note that this approach does not describe collective effects like Thomas-Fermi screening. It is therefore not suitable for an accurate determination of the energy dependence of $\ell$ that requires to take into account the density dependence of both $\xi$ and $\delta V=\hbar v / \xi \sqrt{g/2\pi}$ \cite{Hwang2008, Culcer2010}. Nevertheless, the aim of this work is to show that the transport properties of quasi-ballistic systems are dominated by the interface between the contact and the nanowire and not by the disorder in the nanowire, such that the exact energy dependence of $\ell$ does not play a role for our conclusions as long as $\ell_\text{m} \gtrsim W$. 

The values of $\ell$ explain the clear ballistic features (Fabry-Pérot) observed for $g=0.02$ (see Fig.~\ref{fig2}a of the main text) since $\ell_\text{m}$ lies above the ballistic limit whereas those resonances are averaged out for $g=0.2$ (see Fig.~\ref{fig2}b of the main text), for which $\ell_\text{m}$ is below this limit. Nevertheless, the dips due to intermode scattering for $g=0.2$ reveal the quantum confinement as expected since $\ell_\text{m}$ is above the diffusive limit for any $\varepsilon$. Finally, dips are smeared out by disorder broadening for $g=1$ as shown in Fig.~\ref{fig2}c, pointing to a diffusive regime as expected from $\ell_\text{m}$ that clearly lies below the diffusive limit in a broad range of energy. 

\section{Fit of the transmissions and dependence of the $\gamma$ parameters}
\label{GammaParameters}

\begin{figure}[t]
	\centering
	\includegraphics[width=\columnwidth]{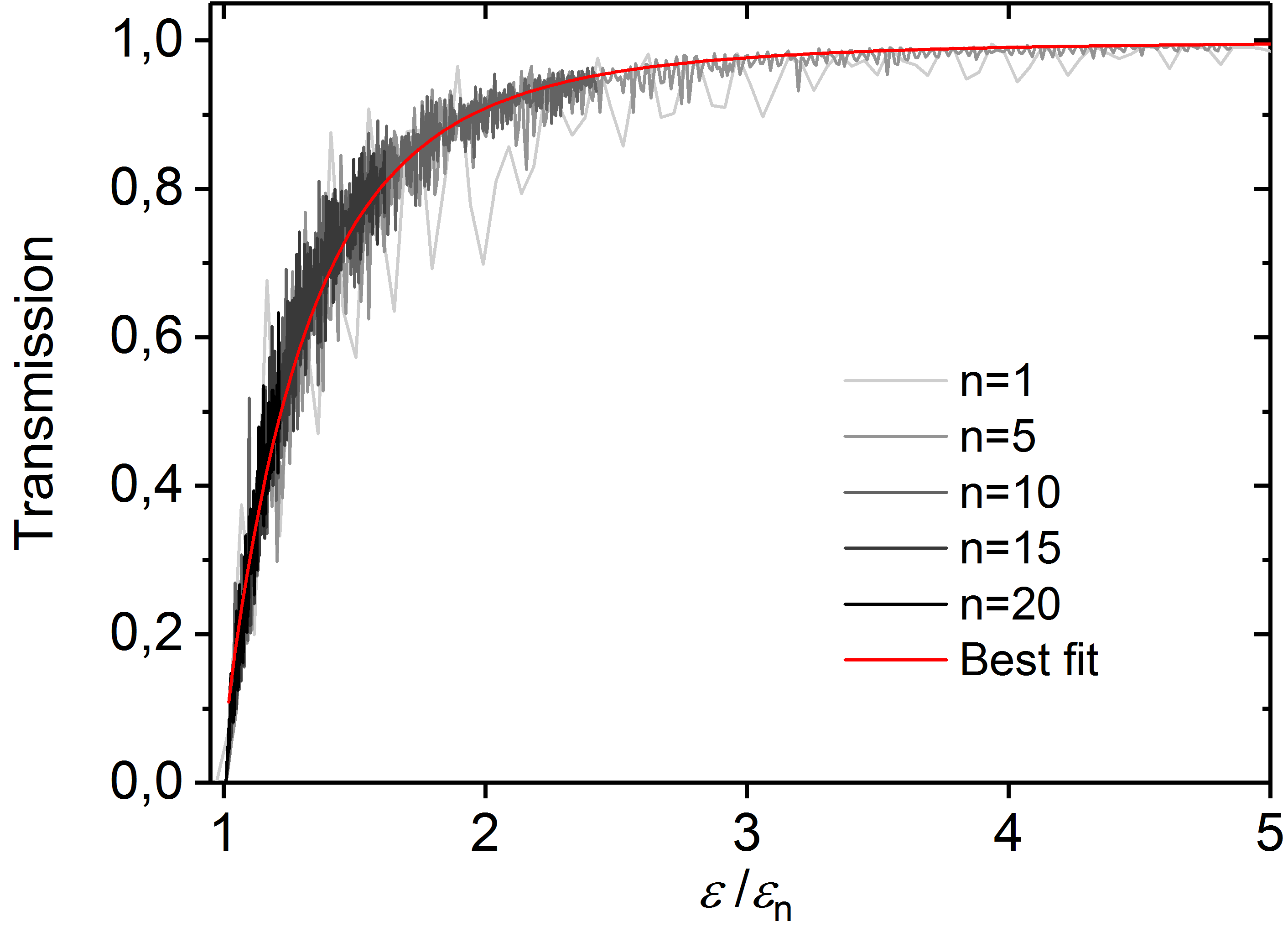}
	\caption[width=\columnwidth]{The transmission for the modes corresponding to $n=1$, $5$, $10$, $15$ and $20$ (from light gray to black) as a function of $\varepsilon_\text{n}/\varepsilon$ and for $g=0.02$ and $L/W=2.5$. The best fit corresponds to the red line.}
	\label{SI-fig2a}
\end{figure}

In order to fit the transmissions and to find the best $\gamma$ parameters, we used the fact that for a weak enough disorder, the transmissions of the different transverse modes depend on the ratio $\varepsilon_\text{n}/\varepsilon$ only. We plot the transmissions of the modes $n=1$, $5$, $10$, $15$ and $20$ as a functions of $\varepsilon_\text{n}/\varepsilon$ in Fig. \ref{SI-fig2a}. We restrict our analysis to five modes only but no significant differences are obtained if all modes are taken into account. The transmissions are roughly superimposed and we fit the data with a single set of free parameters $\gamma_0$ and $\gamma_1$ that describe the quantum corrections for all transmissions. For systems close to the diffusive limit, we adjust the parameters to minimize the error in the high energy limit only, without applying the full fitting procedure. Agreement is found to be excellent over the full energy range for $g<0.1$ (ballistic regime). At high energy ($\varepsilon/\varepsilon_n\gg 1$) excellent agreement can be found for $g$ up to $0.5$.

\begin{figure}[t]
	\centering
	\includegraphics[width=0.9\columnwidth]{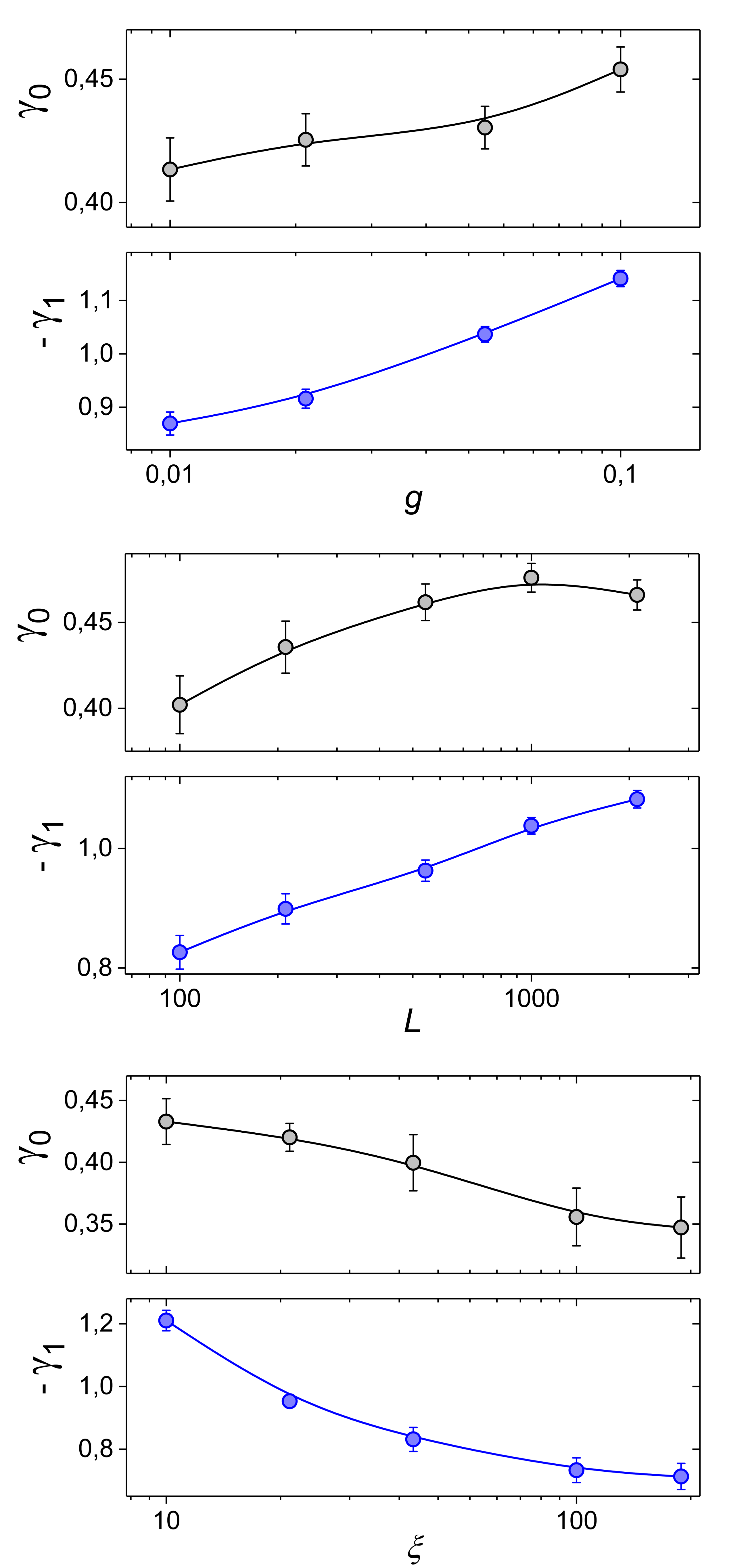}
	\caption[width=\columnwidth]{The dependence of $\gamma_0$ and $\gamma_1$ is shown as a function of $L$, $g$ and $\xi$ in a semi-logarithmic graphs.}
	\label{SI-fig2b}
\end{figure}
The dependence of the $\gamma$ parameters on the disorder strength, wire length and correlation length is weak (see Fig.~\ref{SI-fig2b}, which is plotted on a semi-logarithmic scale). Nevertheless, the dependence of the parameters corresponds to what is roughly expected. The longer the nanowire is, the stronger is the interaction with the disorder and the stronger should be the quantum corrections as reported by the dependence with $L$. Similar evolution is expected and observed for the $g$ dependence. As explained in the main text, the $\xi$ dependence leads to corrections that vanish for long $\xi$ since the $\textbf{q}_2$ scattering process exponentially decays with $\xi$.

Remarkably, the ratio $\gamma_0/\gamma_1$ is almost constant for any value of the different parameters and we have $\gamma_1/\gamma_0\simeq -2$. Hence, the contribution of the two first order transmission corrections to the transport properties compensate each other (see Eq.~\eqref{Conductance} and~\eqref{FanoFactor} in the main text) and the conductance as well as the Fano factor are very close to the non-disordered limit.

\section{Length dependence of the transport properties}
\label{LengthDependence}

\begin{figure*}[!h]
	\centering
	\includegraphics[width=1.4\columnwidth]{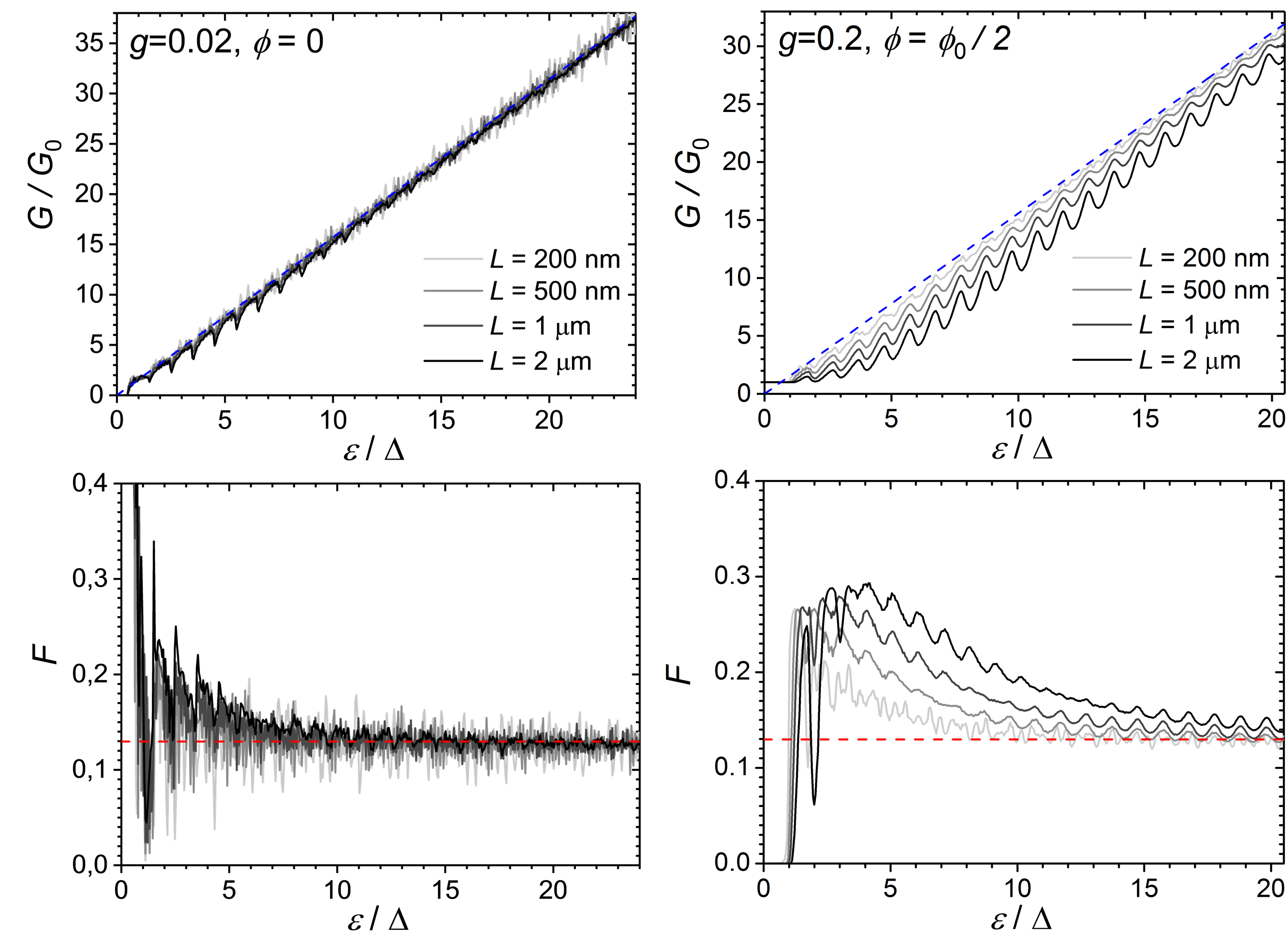}
	\caption[width=\columnwidth]{Left: conductance and Fano factor ($g=0.02$, $W=200nm$, $\xi=10nm$) for different wire lengths $L$. The large number of modes limit is indicated by the red dashed line. Right: conductance and Fano factor ($g=0.2$, $W=200nm$, $\xi=10nm$) for different length. The large number of mode limit is indicated by the blue (conductance) and red (Fano factor) dashed lines.}
	\label{SI-fig3}
\end{figure*}

We focus here on the length dependence of the transport properties ($G$ and $F$) for $g=0.02$ and $g=0.2$ (Fig.~\ref{SI-fig3}).
In the weak disorder limit ($g=0.02$ in Fig.~\ref{SI-fig3} left), the length has only a marginal effect on the conductance or the Fano factor. Generally, it induces sharp dips in the transmission (and then in the conductance) for each energy corresponding to the onset of a transverse mode that does not influence significantly the general transport properties. This is expected since for such a weak disorder, the system is in the ballistic regime ($\ell_\text{m}>2L$) even for $L=2$ $\mu$m (see Fig.~\ref{SI-fig1}).

For a stronger disorder ($g=0.2$ in Fig.~\ref{SI-fig3} right), the system is far below the ballistic limit for $L=500$ nm to $L=2$ $\mu$m and even very close to this limit for $L=200$ nm. We indeed observe a length dependence of the conductance but this dependence is much weaker than in the diffusive regime, for which $G\propto 1/L$. Likewise, the Fano factor remains well below the diffusive limit($F=1/3$), which confirms the ballistic nature of the transport even for large ratio $L/W$ as long as the system is quantum confined. 

\section{Aharonov-Bohm oscillations}
\label{app:AB}

The Fig.~\ref{fig4} in the main text shows the expected shape of the Aharonov-Bohm oscillations for weakly disordered system where the magnetic flux dependence of $\delta G$ is plotted for different positions of the Fermi energy. It indicates a rich content in harmonics of $\delta G$ as observed in experiments \cite{Dufouleur2013,Hong2014}. At low energy ($\varepsilon \lesssim \Delta/2$), the conductance exhibits sharp peaks for $\phi=\phi_0 / 2$ and the Fourier transform of the magnetoconductance contents therefore many harmonics. The harmonic content is strongly reduced when the energy increases but the ratio between the fundamental and the first harmonic remains generally energy dependent. Hence, this ratio is small for $2\varepsilon / \Delta \in \mathbb{N}$ whereas $2\varepsilon / \Delta +1/2 \in \mathbb{N}$ is associated to a large resurgence of the second harmonic. It should be noticed that the presence of intermode scattering is expected to significantly modify the amplitude and the shape of the Aharonov-Bohm oscillations (less harmonics content) for rather strong disorder ($g> 0.1$). 

\bibliographystyle{apsrev4-1}
%

\end{document}